\documentstyle[usp,11pt,epsfig]{article}
\newcommand {\bbrnnumber} {327}
\newcommand {\tnnumber} {09-002}

\def\BaBar{\mbox{\Hbabar}}

\newif\ifdimspec
\def\figsize#1{\dimspecfalse \checkdim#1\end
\ifdimspec
  \def\figureWidth{#1}%
\else
  \def\figureWidth{#1 in}\fi}
\def\checkdim#1{\ifx#1\end \let\next=\relax
  \else \ifcat#1a \dimspectrue \fi \let\next=\checkdim\fi \next}
\newcommand{\lblcaption  }[2]{\caption{#2\label{#1}}}
\newcommand{\Figure }[3]{\begin{figure}[tb]
  \begin{center}
  \mbox{\figsize{#3}\epsfig{file=\chapterdir/#1.eps,width=\figureWidth}}
  \end{center}
  \lblcaption{#1}{#2}
  \end{figure}}
\newcommand{\twoAngFiguresEPS}[6]
{
\begin{figure}[tb]
\begin{center}
\figsize{#4}
\begin{minipage}[t]{3.2in}
\begin{center}
\epsfig{file=\chapterdir/#1.eps,width=\figureWidth,angle=#5}
\end{center}
\end{minipage}
\begin{minipage}[t]{3.2in}
\begin{center}
\epsfig{file=\chapterdir/#2.eps,width=\figureWidth,angle=#6}
\end{minipage}
\end{center}
\lblcaption{#1}{#3}
\end{figure}
}
\newcommand{\twocapFiguresEPS}[5]
{
\begin{figure}[tb]
\begin{center}
\figsize{#5}
\begin{minipage}[t]{3.0in}
\begin{center}
\epsfig{file=\chapterdir/#1.eps,width=\figureWidth}
\end{center}
\lblcaption{#1}{#2}
\end{minipage}
\hfill
\begin{minipage}[t]{3.0in}
\begin{center}
\epsfig{file=\chapterdir/#3.eps,width=\figureWidth}
\end{center}
\lblcaption{#3}{#4}
\end{minipage}
\end{center}
\end{figure}
}
\newcommand{\FigurePS}[3]{\begin{figure}[tb]
  \begin{center}
  \mbox{\figsize{#3}\epsfig{file=\chapterdir/#1.ps,width=\figureWidth}}
  \end{center}
  \lblcaption{#1}{#2}
  \end{figure}}
\newcommand{\twocapFiguresPS}[5]
{
\begin{figure}[tb]
\begin{center}
\figsize{#5}
\begin{minipage}[t]{3.0in}
\begin{center}
\epsfig{file=\chapterdir/#1.ps,width=\figureWidth}
\end{center}
\lblcaption{#1}{#2}
\end{minipage}
\hskip 0.35in
\begin{minipage}[t]{3.00in}
\begin{center}
\epsfig{file=\chapterdir/#3.ps,width=\figureWidth}
\end{center}
\lblcaption{#3}{#4}
\end{minipage}
\end{center}
\end{figure}
}
\def\ifmath#1{\relax\ifmmode#1\else$#1$\fi}

\newcommand{\comment}[1]{}

\def\etal{{\it et al.,}}

\def\ie{{\it i.e.,}}
\def\eg{{\it e.g.,}}

\def\Hbabar{\mbox{{\Huge\bf B}\hspace{-0.1em}{\LARGE\bf A}\hspace{-0.1em}{\Huge\bf B}\hspace{-0.1em}{\LARGE\bf A\hspace{-0.1em}R}}}

\def\babar{\mbox{\sl B\hspace{-0.4em} {\scriptsize\sl A}\hspace{-0.4em} B\hspace{-0.4em} {\scriptsize\sl A\hspace{-0.1em}R}}}

\newcommand{\babarnote}[1] {\mbox{\babar\ {\sl Note}~\# #1}}

\def\TDR{{\sl Technical Design Report}}

\newcommand{\chapterdir}{ch01_physics}

\def\Y#1S  {\ifmath{\Upsilon\rm(#1S)}}

\def\kev  {\ifmath{\mbox{\,ke\kern -0.08em V}}}
\def\mev  {\ifmath{\mbox{\,Me\kern -0.08em V}}}
\def\gev  {\ifmath{\mbox{\,Ge\kern -0.08em V}}}
\def\gevc {\ifmath{\mbox{\,Ge\kern -0.08em V$\!/c$}}}
\def\mevc {\ifmath{\mbox{\,Me\kern -0.08em V$\!/c$}}}
\def\gevcc{\ifmath{\mbox{\,Ge\kern -0.08em V$\!/c^2$}}}
\def\mevcc{\ifmath{\mbox{\,Me\kern -0.08em V$\!/c^2$}}}

\def\in   {\ifmath{\mbox{\,in}}}

\def\mus  {\ifmath{\,\mu\mbox{s}}}
\def\ns   {\ifmath{\,\mbox{ns}}}

\def\hz	  {\ifmath{\mbox{\,Hz}}}
\def\khz  {\ifmath{\mbox{\,kHz}}}
\def\mhz  {\ifmath{\mbox{\,MHz}}}

\def\muv  	{\ifmath{\mbox{\,$\mu$V}}}

\def\nv  	{\ifmath{\mbox{\,nV}}}

\def\muA        {\ifmath{\mbox{\,$\mu$A}}}
\def\nA         {\ifmath{\mbox{\,nA}}}
\def\pA         {\ifmath{\mbox{\,pA}}}
\def\fA         {\ifmath{\mbox{\,fA}}}

\def\pf         {\ifmath{\mbox{\,pF}}}

\makeatletter
\def\@versim#1#2{\vcenter{\offinterlineskip
        \ialign{$\m@th#1\hfil##\hfil$\crcr#2\crcr\sim\crcr } }}
\def\gsim{\mathrel{\mathpalette\@versim>}}
\def\lsim{\mathrel{\mathpalette\@versim<}}
\makeatother

\newcommand{\CsI}{\rm CsI(Tl)}

\newcommand{\Level}[1]	{Level~#1}

\def\Lt		{\Level{2}}
\def\Lh		{\Level{3}}

\def\L\Lt\hreq      {\Lh\ request}

\def\r    {\ifmath{\rm r}}
\def\d    {\ifmath{\bf d}}
\def\a    {\ifmath{\bf a}}
\def\t0	{\ifmath{t_0}}	%
\def\ts	{\ifmath{\tau_s}}	%
\def\tn	{\ifmath{\tau_n}}	%
\def\tb	{\ifmath{\tau_b}}	%
\def\tx	{\ifmath{\tau_x}}	%
\def\tp	{\ifmath{\tau_+}}	%
\def\tm	{\ifmath{\tau_-}}	%
\def\tsb  {\ifmath{\tau_{sb}}}	%
\def\ws	  {\ifmath{\omega_s}}	%
\def\wn	  {\ifmath{\omega_n}}	%
\def\wb	  {\ifmath{\omega_b}}	%
\def\wx	  {\ifmath{\omega_x}}	%
\def\wp	  {\ifmath{\omega_+}}	%
\def\wm	  {\ifmath{\omega_-}}	%
\def\wsb  {\ifmath{\omega_{sb}}}%
\def\wc	  {\ifmath{\omega_c}}	%
\def\wwd	  {\ifmath{\omega_d}}	%
\def\fn	  {\ifmath{f_n}}	%
\def\fd	  {\ifmath{f_d}}	%
\def\fx	  {\ifmath{f_x}}	%
\def\fobs {\ifmath{f_{obs}}}	%
\def\fc	  {\ifmath{f_c}}	%
\def\Td	  {\ifmath{T_d}}	%
\def\is	  {\ifmath{i_s}}	%
\def\isf  {\ifmath{i_{sf}}}	%
\def\in	  {\ifmath{I_n}}	%
\def\id	  {\ifmath{I_d}}	%
\def\ib	  {\ifmath{I_b}}	%
\def\iq	  {\ifmath{I_q}}	%
\def\is	  {\ifmath{i_s}}	%
\def\Is	  {\ifmath{I_s}}	%
\def\inf  {\ifmath{I_f}}	%
\def\ibit {\ifmath{i_{bit}}}    %
\def\vn	  {\ifmath{V_n}}	%
\def\vq	  {\ifmath{V_q}}	%
\def\vbit {\ifmath{v_{bit}}}    %
\def\Cs   {\ifmath{C_s}}        %
\def\Ct   {\ifmath{C}}          %
\def\Cf   {\ifmath{C_f}}        %
\def\Ccal {\ifmath{C_{cal}}}    %
\def\Ciss {\ifmath{C_{iss}}}    %
\def\w    {\ifmath{\omega}} 	%
\def\rhz  {\ifmath{\rm\sqrt{\hz}}} %
\def\mrs  {\ifmath{\rm\,M\,radians/s}} %
\def\pC   {\ifmath{\rm\,pC}} 	%
\def\sgn  {\ifmath{\rm signum}}

\renewcommand{\chapterdir}{.}
\begin{document}
\begin{titlepage}
\begin{flushright}
    \large
   SLAC-TN-\tnnumber \\
   \babarnote \bbrnnumber \\
   October 1996
\end{flushright} \bigskip \bigskip \bigskip
\bigskip
\begin{center} \huge \bf
  \newcommand {\BaBarb} {\boldmath$\BaBar$}
  Noise in a Calorimeter \\Readout System\\ Using Periodic Sampling
\end{center} \bigskip
\begin{center} \Large
   Walter~R.~Innes  \\
   SLAC National Accelerator Laboratory
\end{center} \bigskip \bigskip
\bigskip
\bigskip
\bigskip
\bigskip
\begin{abstract}
Fourier transform analysis of the calorimeter noise
problem gives quantitative results
on a) the time-height correlation, b) the effect of background on optimal
shaping and on the ENC, c) sampling frequency requirements,
and d) the relation between sampling frequency and the
required quantization error.
\end{abstract}
\bigskip
\bigskip
\end{titlepage}
\newpage
\section{Introduction}
\label{sec:intro}

Noise in calorimeter readout electronics has been treated in some
detail~\cite{Radeka,Haller,Dow}.
However, since each author builds the details of
their particular situation into their analysis, it is worthwhile
to review the subject in light of our own experiment. Most
previous studies assume that a single measurement will be made
at a well determined time. Furthermore they assume that filtering
preceding the digitization is limited to relatively simple
analog devices. Since we have chosen to solve our lack of
knowledge of the event time by recording periodic samples,
applying digital signal processing techniques is natural.
This also raises questions about the required sampling rate
and the quantization error that don't occur in the
usual scheme.

Electronics noise and filters are usually analyzed using
Fourier transforms~\cite{Ambrozny,Humphreys,Papoulis}.
In the frequency domain, filtering, differentiation,
integration, and time shifting are all represented by
multiplications. Stochastic noise is most easily
represented by its power spectral density (which is the
Fourier transform of the more complicated auto-correlation function in
the time domain). Many questions such as the signal to noise
ratio for particular parameters and filter design can be
handled entirely in the frequency domain.

\section{The Input Circuit Model}
\label{sec:model}

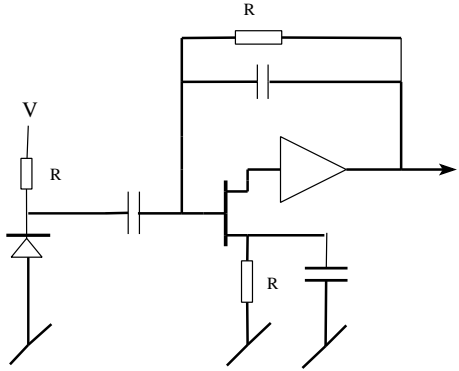
\Figure{circuit}{Simplified photodiode-preamp circuit}{2.5in}

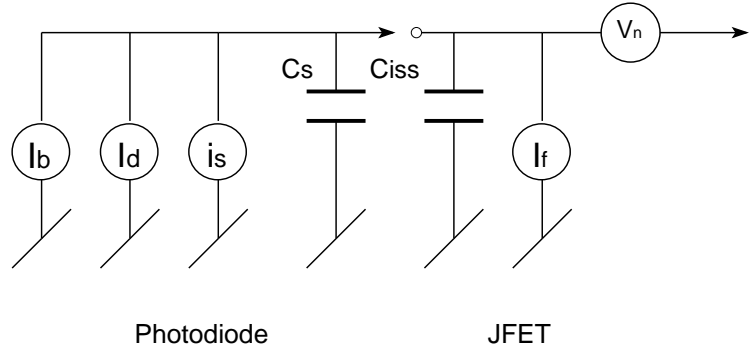
\Figure{circuit_model}{Model for the front end of the calorimeter
readout. \is\ is the current due to the signal and is a function
of time. \id\ is the spectral current noise generator corresponding
to the shot noise of the photodiode. \ib\ is the equivalent current noise
caused by background photons from lost particles.
\inf\ is the FET input current noise, and
\vn\ is the FET input voltage noise. \Cs\ is the capacitance of the
photodiode and \Ciss\ is the common source input capacitance of the FET.}{4in}

Figure~\ref{circuit} shows a simplified version of the calorimeter
input circuitry. Figure~\ref{circuit_model} shows a noise equivalent
representation of the same circuit. The right hand portion represents
the input FET. This is one of many possible equivalent representations
\cite[page 29]{vanderZiel}\cite[page 137]{Ambrozny}.
This particular one has a relatively
simple connection between the physical processes which generate
noise and the elements of the representation. The spectral densities
of the noise generators will correspond to the those given in
manufacturers specification sheets.

\subsection{The photo-diode}

\is\ is a current generator corresponding to the signal generated by
the interactions in the calorimeter. It will be represented by the
function:
\begin{equation}
\is(t) = \frac 1\ts e^{-t/\ts}H(t).
\end{equation}
\ts\ is the decay time of the \CsI\ scintillation and $H(t)$ is the
Heaviside unit step function.
\ts\ is 0.94\mus\ for \CsI.
The Fourier transform of $i(t)$ is
\begin{equation}
\Is(\w)=\int_{-\infty}^\infty{i(t)e^{-j\w t}\,dt} = \frac 1 {1+j\w\ts}\ .
\end{equation}
The spectral power density of the
signal is:
\begin{equation}
\Is^*(\w) \Is(\w) = \frac 1 {1+\w^2\ts^2}.
\end{equation}
For characteristic times represented as \tx, I will use the notation that
the associated radial velocity is $\wx\equiv1/\tx$, the
associated frequency is $\fx\equiv\wx/(2\pi)$, and the associated period
is $T_x\equiv1/\fx$. Using this notation we can write:
\begin{equation}
\Is(\w) = \frac{-j\ws}{\w-j\ws}\ \ {\rm and}\ \ \Is^*(\w)\Is(\w) = \frac{\ws^2}{\w^2+\ws^2}.
\end{equation}
\id\ is the current generator corresponding to the shot noise caused
by the leakage current in the photodiode. This has a ``white''
(independent of \w) spectrum with the value
$\id^2=2 e I_{leak}$ (units of current squared per bandwidth).
The order of magnitude for $I_{leak}$ is
nanoamperes. We'll assume our diode has a leakage of 4\nA.
Since we average two diodes we can use the equivalent
leakage of 2\nA\ for a single diode giving
$\id=25\fA/\rhz$.

\Cs\ is the capacitance of the back-biased photodiode. A typical
value is 80\pf, which we'll use in our estimates.

A photodiode with these properties is the 1N2744.

As much as possible, the discussion will be restricted to the
electronic regime. However, some effects depend on the efficiency
of the conversion of calorimeter shower energy into charge.
When this number is required, I will use the value of
3000 photo-electrons per MeV in each diode.

\subsection{The FET}

\Ciss\ is the FET common source input capacitance. It can range from
a few picofarads to tens of picofarads depending on the choice of FET.
We will return to this choice later.
In the circuit it functions in parallel with \Cs.
Not shown in the figure since they are much less than \Cs\ are
the feedback capacitance \Cf\ and the calibration injection
capacitance \Ccal.
I will call the sum $\Cs+\Ciss+\Cf+\Ccal$ simply \Ct.

\inf\ is a noise generator corresponding the input current noise
of the FET. It is primarily due to the gate leakage current,
although for very large gate areas the effect the thermal noise
of the real part of the input impedance should be checked.
The shot noise acts just like the diode shot noise and the two
leakage currents may be added. In practice the FET gate
leakage current is of the order of picoamperes and is negligible
compared to the diode leakage current. Let
$\in^2=\id^2+\inf^2\approx\id^2$.

\vn\ is a voltage noise generator representing the input voltage
noise of the FET. This is primarily due the thermal noise
in the FET channel. Many authors put a formula,
which shows the dependence of this noise
on the temperature and the forward transconductance, into
their equations~\cite[page 75]{vanderZiel}. I prefer to keep \vn\ as an
explicit term.
The value of \vn\ is usually given on specification sheets and is
also easy to measure.
For junction FETs, \vn\ is approximately independent of
frequency over the range of interest to us. Typical
values are 1 to 2\nv/\rhz.

In order to be more specific I'll take as an example the SANYO
2SK932 JFET.
\Ciss\ is 20\pf\ and \vn\ is $\approx 0.7\nv/\rhz$.
Because
we average the outputs two diode-FET-preamp
combinations on each crystal, we divide
by $\sqrt{2}$ to get 0.5\nv/\rhz.
This is a good FET but not optimally matched to
our problem. We'll return to this issue after we have developed
the signal to noise formulas.

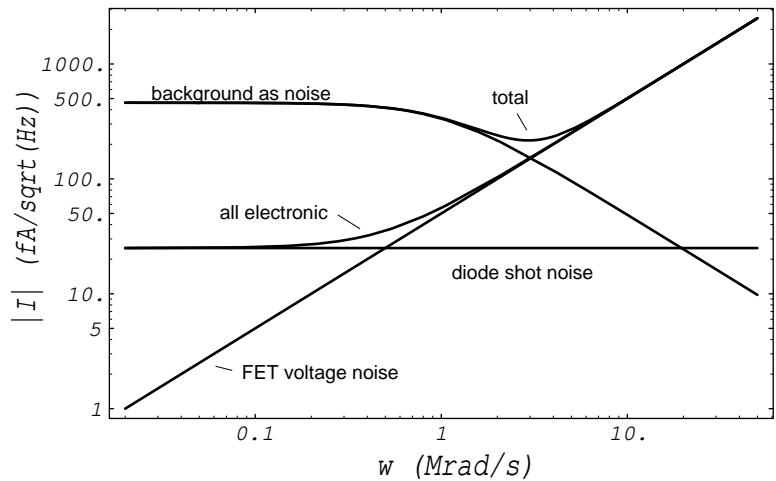
\Figure{noisplt}{The amplitude spectra of the various noise
sources expressed as input current densities. For the
background, nominal conditions are assumed and no energy cutoff
is used. Note that the signal has the same spectrum
as the background noise.}{4in}

\subsection{Background}

\ib\ is a current noise generator representing the sea of photons
generated by particles lost from the beam. This noise resembles
shot noise in that near impulsive elements arrive at random times.
It differs in that the pulse shape is not a delta function but has
the shape of the signal and there is a distribution of pulse areas
instead of just the electron charge.\footnote{In radar theory, noise
with these properties is called ``clutter.''}
If we return to the
derivation of the shot noise formula~\cite[page 81]{Ambrozny},
we see that
the first difference can be treated by giving this noise the
power spectrum of the signal (the Fourier transform of each pulse
is the same as that of the signal
except for phase and magnitude). The second is handled by noting
that the shot noise formula is really a pulse area squared times
a rate: $\id^2 = 2 e^2 f_{leak}$. $e$ is the size of the pulse and
$f_{leak}$ is the mean rate of their occurrence. Therefore we can
get the low frequency limit of the noise spectral density of the
background by doing the integral:
\begin{equation}
\ib^2=\int_0^\infty{2 q^2 B(q)\,dq}
\end{equation}
where $B$ is the rate of background events per pulse area interval
per time interval. The value of this integral can be
inferred from the statistics of histograms such as those
in \babar\ \TDR\ Figure 12-9~\cite{tdr}.
Here I use more recent calculations of the background~\cite{Marsiske}.
Assuming $3000\rm\,pe/\mev/diode$, the result for an average crystal
is $\ib=460\fA/\rhz$ at nominal background.

For convenience
I'll use the symbol \ib\ to represent this low frequency limit
and put the frequency dependence into the formulas explicitly.
Since \in\ and \vn\ are approximately independent of frequency,
I can then treat \ib, \in, and \vn\ as constants.


Note that individual crystals can differ from the average by factors as
large as $4$. In particular, compared an average barrel crystal,
an average endcap crystal has $5/7$th the hit rate,
twice the average energy deposit, and an \ib\ twice as large.

Figure~\ref{noisplt} summarizes all the noise sources.

\subsubsection{Limitations}

The treatment of background as noise is a good description if the
background signal after all processing contains significant contributions
from many background events. If this is not the case, the data
may fall into two classes, those showers little affected by background, and
those with a large background contribution. The former will have
a distribution of measured values consistent with electronic noise,
while the latter may have a wider, highly skewed distribution.
While the background as noise treatment may give a near optimal
procedure for minimizing the rms error, it will widen the
distribution for that class events unaffected by background.
Depending on the physics objective, treating the background
affected events as an inefficiency rather than events with
errors may give the better result.

How well does our case satisfy this many hit condition?
In the background calculation in the preceding paragraph, an average
crystal had $0.11\,\rm hits/crystal/\mus$. Assuming a window of $2\mus$ and
a sum of 25 crystals, this implies 5.5 background hits contributing to the
measurement during nominal conditions. This is enough to give reasonably
Guassian behavior. However, much of \ib\ is generated by the high
energy tail of the background distribution. If we count down from high
energies until get a mean number of hits of at least 2, the cut would
be in the vicinity of $1\mev$. At 10 times nominal this cut would be
close to $3\mev$.

Another consideration is that an approach which
tries to separate a background tail from other noise
is hardly likely to succeed if the magnitude of the background
contribution is of the same size as other errors. The size of the
intrinsic error depends on the size of the signal. For a $100\mev$
shower the expected error is $\approx 2\mev$.
Assuming that background contributions
must be at least $3\mev$ before they can be separated from
noise seems conservative.

If only background hits with energies below $3\mev$ are included,
\ib\ falls to $160\fA/\rhz$.
The background rate falls to $0.10\,\rm hits/crystal/\mus$.

\section{Theorems}

\subsection{$S/N$}

The {\bf Radar Handbook}~\cite[page 4-11]{Skolnik}
reviews the theory of finding
a known signal in noise from the starting point of doing a least
square fit to a finite set of measurements. Under fairly
general conditions this is equivalent to the maximum likelihood
fit and is the best that can be done. The treatment
proceeds by taking the continuous infinite time limit which turns
the usual matrix equations into integral equations. These are then
solved by applying Fourier transforms. For a signal $i(t,a_1,...,a_n)$
with Fourier transform $I(\w,a_1,...,a_n)$
the elements of the inverse of the error matrix for $a_i$
are given by:
\begin{equation}
[{\bf V}^{-1}]_{ik}=\frac 1{2\pi}\int_{-\infty}^\infty
{\left(\left(\frac{\partial I(\w)}{\partial a_i}
\frac{\partial I^*(\w)}{\partial a_k}\right)
/N(\w)\right)d\w} ,
\end{equation}
where $N$ is the double sided noise power spectral density.
This matrix is known as the information matrix
and also as the signal to noise ratio ($S/N$) matrix.

In our case the signal has two parameters, amplitude and time offset:
\begin{equation}
i(t)=a i_0(t-t_0).
\end{equation} This has the Fourier transform
\begin{equation}
I(\w)=a I_0(\w) e^{-j\w t_0}.
\end{equation}
If we define $I_0$ such that the actual value of $a$ is 1 we find
\begin{eqnarray}
\frac{\partial I}{\partial a} &=& I(\w) \mbox{,\rm\hspace{1in} and} \\
\frac{\partial I}{\partial t_0} &=& -j\w I(\w).
\end{eqnarray}
Thus the matrix of integrands in the $S/N$ formula is
\begin{equation}
\left[
\begin{array}{cc}
~~~1~~~ & ~~j\w~~ \\
 -j\w   &  \w^2
\end{array}
\right]
\frac{|I(\w)|^2}{N(\w)}
\end{equation}
Because both $|I(\w)|^2$ and $N(\w)$ are even in \w, the integral of the
off-diagonal elements is 0. Since the matrix is diagonal,
inverting it to get the error matrix ${\bf V}$ is trivial.
If $i(t)$ is normalized to unit area, the $aa$ element of ${\bf V}$ is the square of the equivalent
noise charge (ENC) and the $t_0t_0$ element is the square of product of
the pulse height and the time error (ENTC). Since there is no correlation
between these errors,
{\it a priori} knowledge of the time
does not improve the determination of the amplitude.
The final formulas for the signal to noise are:
\begin{equation}
\left[\frac S N\right]_{aa} =
\frac 1{2\pi}\int_{-\infty}^\infty{\frac{|I(\w)|^2}{N(\w)}\,d\w},\\
\label{eq:snaa}
\end{equation}
and
\begin{equation}
\left[\frac S N\right]_{tt} =
\frac 1{2\pi}\int_{-\infty}^\infty{\frac{|I(\w)|^2}{N(\w)}\,\w^2d\w}
\end{equation}

\subsection{Matched filter}

This result for the error in the amplitude is the same that is given in
many texts as the best that can be achieved using an optimal matched
filter~\cite[page 65]{Humphreys}\cite[page 135]{Papoulis}.
Assume there exists a filter such that the output peaks at $t=0$.
The square of that peak value is given by:
\begin{equation}
o^2(0) =
\left |\frac 1{2\pi}\int_{-\infty}^\infty{I(\w)F(\w)\,d\w}\right |^2,
\end{equation}
where $F(\w)$ is the transfer function of the filter. The mean square noise
after filtering is the same at all times and is given by
\begin{equation}
n_o^2= \frac 1{2\pi}\int_{-\infty}^\infty{N(\w)|F(\w)|^2\,d\w}.
\end{equation}
The signal to noise is the ratio of $o^2(0)$ to $n_o^2$.
Multiplying the integrand of the numerator by $\sqrt{N(\w)}/\sqrt{N(\w)}$
and using
Schwarz's inequality proves that the signal to noise is less than
or equal to the $S/N$ derived from the least square fit method
(equation~\ref{eq:snaa}).
Furthermore, inspection of the $S/N$ equation before this
manipulation shows that the equality can be achieved using a
filter with a transfer function of
\begin{equation}
F(\w)\propto\frac{I^*(\w)}{N(\w)}\ .
\end{equation}
From this I conclude that the best possible result can be
achieved with a filter technique and that the required filter is readily
calculated given knowledge of the shape of the signal and
the spectrum of the noise.

The impulse response to the optimal filter is
\begin{equation}
f(t)\propto\frac 1{2\pi}\int_{-\infty}^\infty{\frac{I^*(\w)}{N(\w)}\,e^{j\w t}d\w}\ ,
\end{equation}
and the shape of a noiseless signal after filtering (the convolution
of the signal with impulse response) is
\begin{equation}
o(t)\propto\frac 1{2\pi}\int_{-\infty}^\infty{\frac{|I(\w)|^2}{N(\w)}\,e^{j\w t}d\w} \ .
\end{equation}
A useful normalization for the filter can be achieved by
dividing by the signal to noise ratio (equation~\ref{eq:snaa}).
This yields $o(0)=1$.
In the following discussion, $F$, $f$, and $o$ are so normalized.

\section{Application to our problem}

We have now collected all the necessary input information and
the tools we need to design an optimal filter and calculate the
signal to noise. Before tackling the full problem lets explore
some special cases so that we can get an understanding of
the effect of each input, can compare our results with previous
work, and gain confidence in the method. Some of the formulas
are followed by a bracketed reference to the source of the
evaluation of the previous integral. Actually most of the
integrals are straightforward (but sometimes tedious) to do
using contour integration. This is a consequence of the fact that
the signals, backgrounds, and analog filters all have exponential
shapes in the time domain. More general signal shapes are less
tractable.

\subsection{No background ($\ib=0$) and $\ts=0$}

This is the usual treatment when the ``ballistic deficit''
is ignored.
$\is(t)$ is a delta function,
$\Is(\w)=1$, and $2N(\w)=\in^2+\vn^2\w^2\Ct^2$.

Before starting lets evaluate some useful constants related to
the inputs so that we can use them to evaluate the results as we
go: $\tn\equiv\vn\Ct/\in = 2.0\mus$, $\wn\equiv1/\tn=0.50\mrs$, and
$\fn=80\khz$.
\begin{eqnarray}
\left[\frac S N\right]_{aa} &=&
\frac{\wn^2}{\pi \in^2}\int_{-\infty}^\infty{\frac{1}{\wn^2+\w^2}\,d\w} \\
&=&\frac\wn{\in^2} \\
ENC^2 &=& \in^2\tn = \in\vn\Ct\\
\left[\frac S N\right]_{tt} &=&
\frac{\wn^2}{\pi \in^2}\int_{-\infty}^\infty{\frac{\w^2}{\wn^2+\w^2}\,d\w} \\
&=&\infty \\
ENTC^2 &=& 0\\
F(\w) &=& \frac {2\wn}{\wn^2+\w^2}\\
f(t) &=& e^{-|t|/\tn}\\
o(t) &=& e^{-|t|/\tn}
\end{eqnarray}

Substituting our sample values, we find that $ENC=223\,e$, and the
corner frequency of the optimal filter is 80\khz.

\subsection{No background ($\ib=0$) and $\ts>0$}

This is the case usually treated, ballistic deficit included.
\begin{eqnarray}
\Is(\w)&=&\frac{-j\ws}{\w-j\ws}\\
2N(\w)&=&(\in^2+\vn^2\w^2\Ct^2) = \in^2(\w^2+\wn^2)/\wn^2\\
\left[\frac S N\right]_{aa} &=&
\frac{\ws^2\wn^2}{\pi \in^2}
\int_{-\infty}^\infty{\frac{1}{(\w^2+\ws^2)(\w^2+\wn^2)}\,d\w} \\
&=&\frac{\ws\wn}{\in^2(\ws+\wn)} \mbox{\hspace{1in} \rm[Dwight~856.31]}\\
ENC^2 &=& \in^2(\tn+\ts) = \in\vn\Ct+\in^2\ts\\
\left[\frac S N\right]_{tt} &=& \frac{\ws^2\wn^2}{\pi \in^2}
\int_{-\infty}^\infty{\frac{\w^2}{(\w^2+\ws^2)(\w^2+\wn^2)}\,d\w} \\
ENTC^2 &=& \in^2\tn\ts(\tn+\ts) = ENC^2\tn\ts
\mbox{\hspace{0.5in} \rm[Maple]}\\
F(\w) &=& \frac {2j\wn(\wn+\ws)}{(\w+j\ws)(\w^2+\wn^2)}\\
f(t) &=& \frac{\wn(\wn+\ws)}{\pi}\int_{-\infty}^\infty
{\frac{(\ws+j\w)}{(\w^2+\ws^2)(\w^2+\wn^2)}e^{j\w t}d\w}\\
 &=& \frac{\wn(\wn+\ws)}{(\wn^2-\ws^2)}
\left(e^{-|t|/\ts}-\frac\ws\wn e^{-|t|/\tn}-\sgn(t)
\left(e^{-|t|/\ts}-e^{-|t|/\tn}\right)\right)\\
 &=& \left( e^{-|t|/\tn}+
H(-t)\frac{2\ts}{\tn-\ts}\left(e^{-|t|/\tn}-e^{-|t|/\ts}\right)\right)\\
o(t)&=&\frac{\wn\ws(\wn+\ws)}{\pi}
  \int_ {-\infty}^\infty{\frac{e^{j \w t}}{(\w^2+\ws^2)(\w^2+\wn^2)}\,d\w}\\
  &=&\frac 1{(\tn-\ts)}
  \left(\tn e^{-|t|/\tn}-\ts e^{-|t|/\ts}\right)
\end{eqnarray}

Substituting our sample values, we find that $ENC=239\,e$,
$ENTC=327\mus\,e$ (1.1\ns\ for a 100\mev\ deposit), and the
corner frequency of the optimal filter is still 80\khz.
The ENC corresponds to and equivalent noise energy (ENE) of
80\kev.

\subsection{$\ib>0$, $\vn=0$, and $\in=0$}

This special case of no electronic noise
is explored to give us confidence that the
filter technique does separate the desired signal from the
background noise.
\begin{eqnarray}
\Is(\w)&=&\frac{-j\ws}{\w-j\ws}\\
2N(\w)&=&\frac{\ws^2\ib^2}{\ws^2+\w^2}\\
\frac{|\Is|^2}{N}&=&\frac2{\ib^2}\\
\left[\frac S N\right]_{aa} &=& \infty\ ,\ ENC^2=0\\
\left[\frac S N\right]_{tt} &=& \infty\ ,\ ENCT^2=0\\
f(t) &\propto& \frac1{\pi}\int_{-\infty}^\infty
{(\ws-j\w)e^{j\w t}d\w}\\
&\propto&2\left(\delta(t)-\ts\delta'(t)\right)\\
o(t)&\propto&\frac1{\pi}\int_{-\infty}^\infty{e^{j\w t}d\w}
\,\,\propto\,\,2\delta(t)
\end{eqnarray}

Because the signal and (background) noise have the same spectrum,
the signal to
noise ratio is independent of frequency. This results in infinite
signal to noise integrals and no error on the pulse height and
time determinations. In the absence of other noise, the matched
filter turns both the signal and background events into
delta functions which are always separable.

\subsection{The general case}

Now we will treat our general case, with all noise and width processes
active. The signal is as for the previous two cases. The noise
is the sum of the noise in those cases.

\begin{eqnarray}
\Is(\w)&=&\frac{-j\ws}{\w-j\ws}\\
|\Is(\w)|^2&=&\frac{\ws^2}{\ws^2+\w^2}\\
2N(\w)&=&\frac{\in^2}{\wn^2}(\wn^2+\w^2)+\frac{\ws^2\ib^2}{\ws^2+\w^2}
\label{eq:genNin}\\
\left[\frac S N\right]_{aa} &=&
\frac{\ws^2\wn^2}{\pi\in^2}\int_{-\infty}^\infty
{\frac{1}{\w^4+(\ws^2+\wn^2)\w^2+\wn^2\ws^2(1+\ib^2/\in^2)}\,d\w} \\
\label{eq:genN}
b^2&\equiv&1+\ib^2/\in^2
\mbox{\hspace{.5in}\rm and\hspace{.5in} }\wb\equiv\wn b\\
\left[\frac S N\right]_{aa}
&=&\frac{\ws\wb}{(\in^2+\ib^2)\sqrt{2\wb\ws+(\wn^2+\ws^2)}}
 \mbox{\hspace{1in} \rm[Dwight~857.11]}\\
\label{eq:optsn}
ENC^2 &=& (\in^2+\ib^2)\sqrt{\tb\ts}\sqrt{2+\tb/\ts+\ts/(b^2\tb)}\\
\left[\frac S N\right]_{tt} &=&
\frac{\ws^2\wn^2}{\pi\in^2}\int_{-\infty}^\infty
{\frac{\w^2}{\w^4+(\ws^2+\wn^2)\w^2+\wn^2\ws^2(1+\ib^2/\in^2)}\,d\w} \\
ENTC^2 &=& \in^2\tn\ts\sqrt{\tn^2+\ts^2+2b\tn\ts)} = ENC^2\tb\ts
\mbox{\hspace{.35in}\rm[Prudnikov 2.2.10 \#4 page 313]}\hfil\\
F(\w) &=&
ENC^2\,\,\frac{2\ws\wn^2(\ws+j\w)}{\in^2(\w^4+(\ws^2+\wn^2)\w^2+\ws^2\wn^2 b^2)}
\\
f(t) &=& ENC^2\,\,\frac{\ws\wn^2}{\pi\in^2}\int_{-\infty}^\infty
{\frac{(\ws+j\w)e^{j\w t}}
{(\w^4+(\ws^2+\wn^2)\w^2+\ws^2\wn^2b^2)}d\w}\\
 &=& ENC^2\,\,\frac{2\ws\wn^2}{\pi\in^2}
\left(
\int_0^\infty{\frac{\ws\cos(\w t)-\w\sin(\w t)}
{(\w^4+(\ws^2+\wn^2)\w^2+\ws^2\wn^2b^2)}d\w}\right)\\
\wp&\equiv&\sqrt{(\wb\ws+(\ws^2+\wn^2)/2)/2}\\
\wm&\equiv&\sqrt{(\wb\ws-(\ws^2+\wn^2)/2)/2}\\
f(t) &=& ENC^2\,\,\frac{\wn\ws}{2\in^2}e^{-\wp|t|}\left(
    \frac1{b\wp}\cos(\wm t)+\frac1{b\wm}\sin(\wm|t|)-
    \frac{\wn}{\wp\wm}\sin(\wm t)\right)\\
&&\nonumber\mbox{\hspace{2.5in}\rm[Prudnikov 2.5.10 \#15 and\#17 page 397]}\\
    &=&e^{-|t|/\tp}\left(\cos(t/\tm)+
    \left(\frac\tm\tp\sgn(t)-\frac\tp\tb\right)\sin(t/\tm)
    \right)
     \\
o(t)&=&ENC^2\,\frac{\wn^2\ws^2}{\pi\in^2}\int_{-\infty}^\infty{
\frac{e^{j \w t}}{(\w^4+(\ws^2+\wn^2)\w^2+\ws^2\wn^2b^2)}\,d\w} \\
&=&e^{-|t|/\tp}\left(
\cos(t/\tm)+\frac\tm\tp\sin(|t|/\tm)\right)
\end{eqnarray}

The filter output for a noiseless signal pulse is shown
in Figure~\ref{ot}.

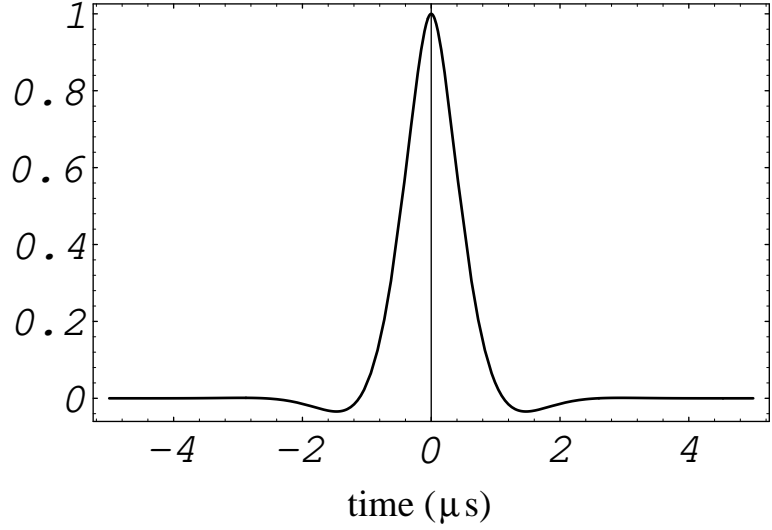
\FigurePS{ot}{A noiseless signal after passing through the optimal
filter. The peak gives the estimate for the signal charge and
time. In this case t=0 and the charge is unity.}{4.0in}

\subsection{$\in = 0$, $\ws<<\wb$}

The results in the previous section are a bit complex, but can
be simplified for our case if we note that \ib\ is more
than 10 times \in\ even at nominal background levels.
In addition, \wb\ is more than five times \ws.
Setting $\in=0$, using the approximation
$\wb >>\ws$, and defining
\mbox{$\wsb\equiv\sqrt{\wb\ws/2}\approx\wm\approx\wp$},
we find:

\begin{eqnarray}
\Is(\w)&=&\frac{-j\ws}{\w-j\ws}\\
|\Is(\w)|^2&=&\frac{\ws^2}{\w^2+\ws^2}\\
2N(\w)&\approx&\frac{\ib^2(\w^4+4\wsb^4)}{\wb^2(\w^2+\ws^2)}\\
|\Is(\w)|^2/N(\w)&\approx&\frac{8\wsb^4}{\ib^2(\w^4+4\wsb^4)}\\
ENC^2&\approx&\ib^2\tsb
=\sqrt{2\ts\ib^3\vn\Ct}\\
ENTC^2 &\approx& ENC^2\ts\tb
=\sqrt{2\ib\ts^3\vn^3\Ct^3}\\
F(\w) &\approx& \frac{j\wb\wsb(\w-j\ws)}{(\w^4+4\wsb^4)}\\
f(t)&\approx&
e^{-|t|/\tsb}\left(
\cos(t/\tsb)+\sin(|t|/\tsb)-\sqrt{2\ts/\tb}\sin(t/\tsb)\right)\\
o(t)&\approx&
e^{-|t|/\tsb}\left(\cos(t/\tsb)+\sin(|t|/\tsb)\right)
\end{eqnarray}

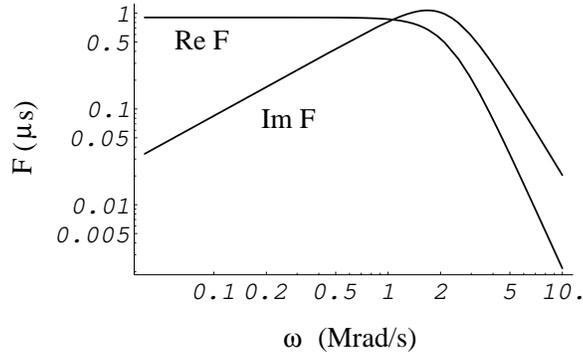
\FigurePS{Fw}{The optimal filter in the frequency domain.}{3. in}
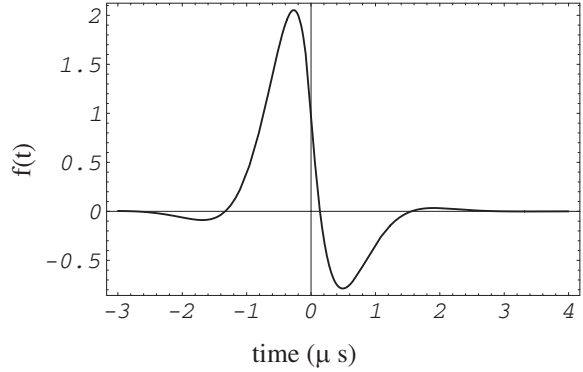
\Figure{ft}{The optimal filter in the time domain.}{3. in}

The optimal filter is shown in the frequency and time domains in
Figures~\ref{Fw} and~\ref{ft}.

Now for some values: $\tb=\vn\Ct/\ib=0.11\mus$,
the filter corner $\sqrt{\ts\tb}=0.32\mus$, $ENC=1930$\,e, and
$ENTC=620$\mus\,e or 2.1\ns\ for a 100\mev\ deposit.
The frequency corner of the optimum filter is
0.50\mhz.

For our worst case of times $10\times$nominal background,
\ib\ goes up $\times3.1$,
\tb\ goes down $\times3.1$, and $\sqrt{\ts\tb}$ goes down $\times1.78$ to 0.17\mus.
$ENC=4460\,e$, $ENTC=760\mus\,e$, and the corner frequency goes
to $0.90\mhz$.
If the $3\mev$ cut were used to calculate \ib, \ib\ would be lower
by a factor of $2.4$ and the values for $\times10$ would be about
the same as the uncut nominal values.

\section{Picking the best FET}

We see that the $ENC$ depends on the FET properties only in
the combination \vn(\Ciss+\Cs). The smaller this value the
better. For a given FET design, these parameters can be varied
by changing the gate area $A$ (or equivalently connecting FETs
in parallel). \vn\ decreases as $1/\sqrt{A}$ while \Ciss\ is
proportional to $A$. \vn\Ct\ is a minimum with respect to $A$
when $\Ciss=\Cs$. This implies that our sample FET should be scaled to 4 times
its original gate area. Making this change would decrease \vn\Ct\ by
a factor of 1.25. Since $ENC$ goes as the fourth root of \vn\Ct,
$ENC$ would decrease only 5\%. On the other hand, $ENTC$ would
decrease by 16\%.

\section{Sampling Requirements}

So far we have assumed that the signals are ideally and continuously
measured. In fact they will be sampled with limited precision. What
restrictions on the sampling rate, length, and accuracy will
prevent the loss of information?

\subsection{Sampling frequency}
\label{samfreq}

The sampling frequency must be high enough to
capture all frequencies with useful $S/N$. It must also be high
enough to avoid misrepresenting noise at higher frequencies as
noise in the signal region (aliasing).
On the other hand the sampling frequency should be no higher than necessary.

What is the highest useful frequency? We note that $S/N$ is
approximately constant up to $\w=\wsb$ and then decreases as
$1/\w^4$. If we throw away all information above $\w=\wc$ the
fraction of information lost is approximately $(3/4)(\wsb/\wc)^3$.
For a 20\% loss of information this implies that $\wc\ge1.5\wsb$, and
that $\fc>0.6\mhz$. The fractional loss in time signal noise
is somewhat larger, being given approximately by $(3/4)(\wsb/\wc)$.
Thus our choice based on the amplitude error will lose us half
of our time information and increase our time error by 25\%. This
is probably acceptable. The exact formula for the fractional
loss of amplitude information is
\begin{equation}
1-\frac1\pi\left(\frac1 2\log\left(
\frac{\wc^2+2\wsb\wc+2\wsb^2}{\wc^2-2\wsb\wc+2\wsb^2}\right)
+\tan^{-1}\left(\frac{2\wc\wsb}{(2\wsb^2-\wc^2)}\right)\right)
\end{equation}

Assume that we have one stage of near ideal integration before the
digitizer (the charge sensitive preamp)
followed by a near ideal differentiator. This pair gives a
band pass filter with a pass band from
a low frequency to \fc. The fall in the
stop band is $1/\w$. Further assume that there
is an additional two pole low pass filter with a corner
at \fc\ (drop off like $(\wc/\w)^2$ for $\w>\wc$)
between the preamp and the digitizer.
If we sample at frequency
\fd, noise at frequencies above $\fd/2$ will appear at
$\fobs=\fd-f$.
We are interested in studying the contribution to $N$
from $\vn^2$ which drops off as
$(\wc/w)^4$ after all the analog filters.
We have sufficient protection against aliasing if
$\vn^2$ at $\fd/2-\fc\ll\vn^2$ at \fc.
This is satisfied if $(\fc/(\fd/2-\fc))^4 < 4$.
This implies $\fd > 3.4\fc$,
which is $3.4(1.5)/(2\pi\sqrt{\ts\tb}) = 0.8/\sqrt{\ts\tb}$.
This is 2.5\mhz\ at nominal
background and 4.7\mhz\ at $\times 10$ nominal. 3.7\mhz\ is adequate adequate
for nominal backgrounds and marginal for $\times 10$.

\subsection{Sample length}

The number of samples required is related to the lowest frequency of
interest. So far we have assumed that all integrals go from
zero frequency. Pile up makes this impractical as does the fact that
we can't use all past history and wait forever to get an answer.

Lets look at the pile up problem first.
Allowing for variations from
crystal to crystal, and warm diodes, we should plan for at least
20\nA\ of total current in a diode.
If the offset from this current is to stay within
10\% of full scale, the product of this current and the
integration time must be less than 10\% of the
charge of a 14\gev\ shower, \ie\ $<0.5\pC$. This limits the
integration time to 200\mus. This sets the minimum value for the
frequency corner of the first integrator.

This gives 740 samples as the maximum useful number
at 3.7\mhz. This is a large number.
What is the smallest number of samples we can use
without significant loss of information?  For the amplitude parameter,
the $S/N$ is flat up to the corner frequency and then drops
rapidly. For our lowest noise case ($\ib=0$ during
source calibration for example), we have a corner
frequency of 80\khz. If we permit a 20\% loss of
information, we can cut off the integral
at the low frequency side at 13\khz. This is less than three times the
previous integration limit leaving us little choice and many
samples. The problem here is that the sample rate is much higher than needed
for this case. Once filtering has been performed,
the sample set could be decimated, but most of the work is done by then.
In the nominal background case the corner frequency is 400\khz, and
the integral could be cut off at 80\khz\ for a sample length
of 12.5\mus\ or 46 samples. This would be the normal operating mode.
If the sampling frequency were optimized for this nominal
background level, the number of samples would be less than 25.

\subsection{Sampling summary}

We can summarize the conclusions of the last two sections by noting that
the number samples that must be dealt with is proportional to the ratio
of the highest and lowest frequencies of used.
The proportionality constant lies
somewhere between 2.5 and 5 depending of the details of the system
and how ``highest'' is defined.
If the signal to noise ratio is approximately constant from low
frequencies up to the highest frequency of interest then the
fraction of the available information retained (the efficiency)
of the system is given by $1-f_{lowest}/f_{highest}$. An
interesting but not necessarily relevant observation:
the maximum information per sample is obtained when
$f_{lowest}$ is one half of $f_{highest}$ and the efficiency
is 50\%.

\section{Quantization Error}

\subsection{With no net integration or differentiation}

With the same assumptions about filtering
as in the previous section we can estimate the
requirements on the quantization error. We assume there is a low
pass filter such that there is no aliasing, no loss of information,
and there are offsetting integrators and differentiators such
that there is no net effect in the pass band.
Under such conditions, quantization error may be referred back to the
input where it appears as another current noise source.

According to the sampling theorem~\cite[page 141]{Papoulis},
if the anti-aliasing conditions are met, the original signal
may be reconstructed from the samples by the interpolation
formula
\begin{equation}
\is(t)=\sum_{n=-\infty}^\infty
\is(n\Td)\frac{\sin((\wwd/2)(t-n\Td))}{(\wwd/2)(t-n\Td)}
\end{equation}
where \Td\ is the sampling period, and \wwd\ is related to it in the
usual way. An error in a sample can be represented as
a function proportional to one element
of this sum. The Fourier transform of such a function
is flat out to $\fd/2$.
The amplitude of this noise pulse is drawn from a square distribution
whose width is given by the bit resolution (or the effective bit resolution
for a non-ideal ADC). The rms area of the pulse is then $\Td\ibit/\sqrt{12}$.
The rate of pulses with effectively random phase is $\fd\equiv1/\Td$, and the
noise spectral density is flat with a value of $\iq^2=\ibit^2/(12\fd)$.
This is to be added to the FET input current noise. We should compare
this to $\in^2$
for very quiet conditions, and to $\ib^2$ for nominal conditions.

First lets treat the quiet case
for which $ENC^2$ goes as
$\sqrt{\in^2}$.
If we wish
to restrict our error increase to less than 5\% due to this
source we require that $\iq^2<\in^2/5$. Therefore
\ibit\ should be less than $\sqrt{2\fd}\,\in$. For the example
diode and our sampling rate, this is $\ibit<68\pA$.
The current at the signal peak is $q/\ts$ if there is no analog
shaping. For a triple RC filter with a time constant of 0.25\mus,
this is reduced by a factor of 2.
Assuming that 12\gev\ somehow gets into
one crystal, the maximum peak current would be 3\muA. This implies
a dynamic range requirement of 43,000 or 15.4 bits.
Since the signal is not calibrated,
we need to add a bit for gain variations suggesting that
16.5 effective bits of dynamic range are required during quiet conditions.
Note that better light collection imposes a greater dynamic range requirement.

During nominal conditions,
$ENC^2$ goes as $\sqrt{\ib^3}$. The 5\% error increase condition
implies that $\iq^2<\ib^2/7$, and that
\ibit\ be equivalent to less than $\sqrt{1.7\fd}\,\ib$. For nominal
background conditions and our sampling rate this is $\ibit<710\pA$.
The dynamic range requirement is 4,200 or 12 bits. With an extra
bit for calibration differences, this is 13 bits.

In practice, many crystals are summed to measure a shower. Both the
electronics and background noise increase as $\sqrt{m}$, where $m$
is the number of crystals included. The dynamic range required decreases
as $\sqrt{m}$. For a 9 crystal sum and nominal background conditions,
the number of bits required is 10.5.

\subsection{With one net integration}

If we assume that instead of offsetting integrators and differentiators,
we have only one integrator and then the low pass filter, the
quantization error behaves as a voltage noise when referred to the input.
The rms area of the error pulse is $\Td\vbit/\sqrt{12}$, and the
noise spectral density is flat with a value of $\vq^2=\vbit^2/(12\fd)$.
This is to be added to the FET input voltage noise $\vn^2$.
Recall that in our case $ENC^2$ goes as $\sqrt{\vn^2}$
in the quiet case and as $\sqrt[4]{\vn^2}$ in the nominal case. If we wish
to restrict our error increase to less than 5\% due to this
source we require that $\vq^2<\vn^2/10$ in the quiet case and
$\vq^2<\vn^2/5$ in the nominal case. This implies that
\vbit\ be equivalent to less than $\sqrt{1.2\fd}\,\vn$ and
$\sqrt{2.4\fd}\,\vn$, respectively. For the example
FET and our sampling rate this is $\vbit<1.05\muv$ and $\vbit<1.5\muv$,
which given the
source capacitance corresponds to a charges of $<660$ and $<940$ e, and to
a shower energies of $<0.22$ and $<0.31\mev$.
Assuming that 12\gev\ somehow gets into
one crystal, the dynamic ranges required are 55,000 and 42,000,
or 15.7 bits for quiet conditions and 15.4 bits for nominal conditions.
Since the signal is not calibrated,
we need to add a bit for gain variations suggesting that
17 effective bits of dynamic range may be sufficient.

All of the above dynamic range calculations address the sufficient conditions
for which the quantization error will not contribute to the electronic
(and background) noise. They do not treat whether or not this dynamic range is
necessary given the inherent energy fluctuations of the shower process.

The considerations for multi-range digitizing have been addressed
by Al Eisner and Gunther Haller and remain unchanged.
For larger pulses, sources of error other than electronics dominate.

\section{Filter Implementation and Interpolation}

There is still the problem of signal extraction.
Because it is non-causal, a matched filter cannot be implemented
as a simple analog filter.
A procedure for
directly calculating the coefficients of a tapped delay line
approximation to the optimum matched filter is described by
Papoulis~\cite[page 327]{Papoulis}. The result is exactly parallel
to the usual least square fit formulas.
The autocorrelation function, $n_f$, is the inverse Fourier transform of the noise power spectral density after the analog filter.
The elements of \r, the error matrix for the
samples, are given by $n_f(|l-k|T)$.
The expected values \a\ for the samples is given by $\isf((l-m)T+t)$,
where \isf\ is the
expected signal after analog filtering and $t$ is the time
offset of the signal from sample $m$. The coefficients of the tapped delay
line filter are given by $\r^{-1}\a$. The optimally filtered estimate
for the signal at time $t$ is given by $o(t)=\d\r^{-1}\a$, where \d\ is the
set of samples. For the case of the triple RC analog filter, the
coefficients each have four terms with different $t$ dependencies.
The estimate for $o(t)$ also has four such terms and its
derivative with respect to $t$ has three.
Calculating $o(t)$ takes four multiply and adds for each sample used.
Finding the root of the
expression for the derivative with Newton's method
should not take many iterations.
Each iteration involves calculating one exponential and approximately
a dozen multiply and adds. The presence of multiple peaks in the time
window or the absence of any peaks would complicate this last step.
A possibility is to not try to find the peak, but to report the
coefficients of the four terms instead.

Knowing the event time would reduce the time to get an estimate for
$o(t)$ by a factor of four, and there would no need to search for
the maximum.

\section{Less Than the Best}

The previous sections deal largely with optimum solutions, albeit
with some practical limitations. This section will examine the
loss of information if other that optimal filtering is used.
It may be less than optimal in that it is not matched, \eg\
an all analog filter, or in that the pass band
is optimized for another condition.

\twocapFiguresEPS
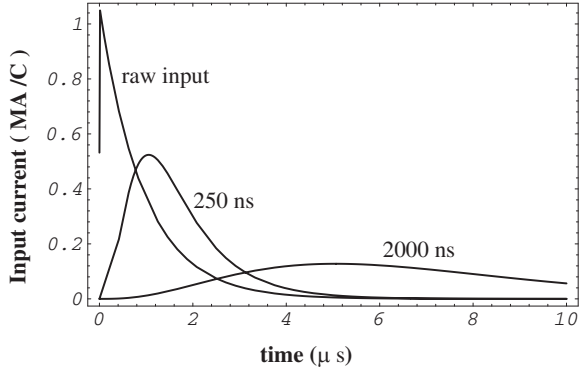
{oof}{Raw input current signal and the signal after
triple RC filters with time constants of $0.25\mus$ and
$2.0\mus$.}
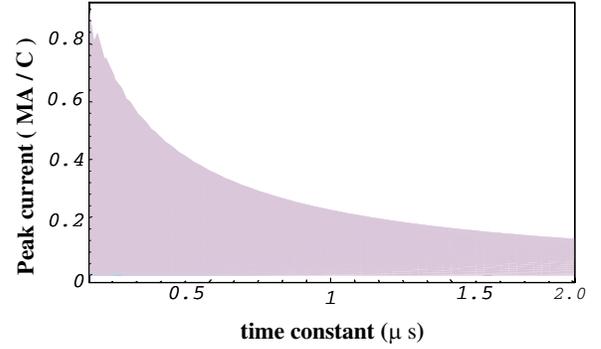
{ooff}{The peak signal current after a triple RC
filter vs. the filter time constant}
{3.0in}

Let us look at the case of an all analog filter. In this case
all filtering is done prior to digitizing and the only
digital processing performed is interpolation. I will take the
\babar\ TDR design (described in subsection~\ref{samfreq}). This
has a charge sensitive preamp followed by a CR and then two RC
filters. Together these give the equivalent of three RC filters,
all with the corner frequency \fc. This has the transfer function
\begin{equation}
F(\w)=\left(\frac{-j\wc}{\w-j\wc}\right)^3 .
\end{equation}
The signal is
\begin{equation}
I(\w)=\left(\frac{-j\ws}{\w-j\ws}\right) .
\end{equation}
\twocapFiguresPS
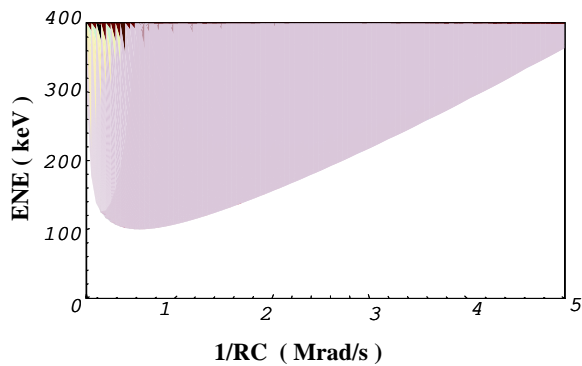
{kev}{The Equivalent Noise Energy (electronic only)
of a single sample at the peak after a triple RC filter versus
the radial velocity of the filter corner ($1/RC$).}
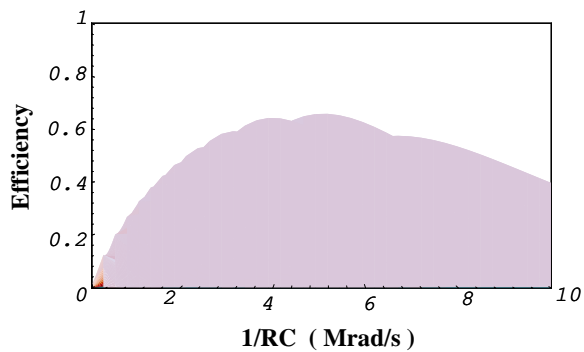
{eff}{The efficiency vs. 1/RC for nominal
background conditions.}
{3.in}
The time dependence of the output signal is given by the
inverse Fourier transform of their product. This integral can
be done by contour integration. If the contour is completed by
a semi-circle at infinity around the upper half plane (for $t>0$), the
integral can be deduced from the residues at the poles
$j\wc$ and $j\ws$. The result is
\begin{equation}
o(t)=\frac{\wc^3\ws}{(\ws-\wc)^3}\left\{\left(\frac{(\ws-\wc)^2}{2}t^2
-(\ws-\wc)t+1\right)e^{-\wc t}-e^{-\ws t}\right\} .
\end{equation}
This is shown for several time constants in figure~\ref{oof}.
The peak value vs. time constant is shown in figure~\ref{ooff}.
Since there are no poles in the lower half plane,
the integral is $0$ for $t<0$.

\twocapFiguresPS
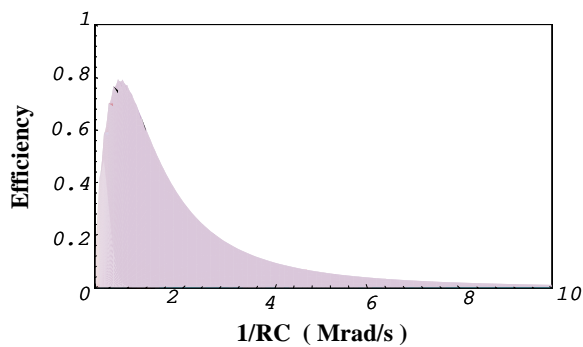
{effa}{The efficiency vs. 1/RC for low
background conditions.}
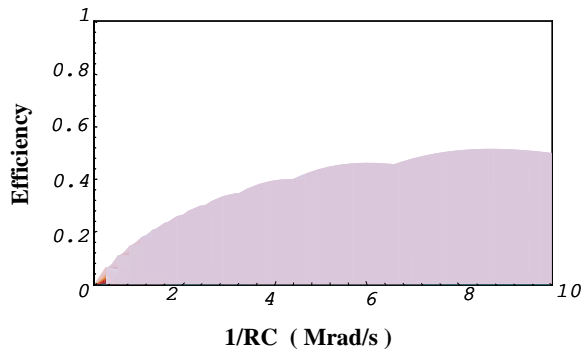
{effb}{The efficiency vs. 1/RC for $10\times$ nominal
background conditions.}
{3in}

The mean square noise is given by the inverse Fourier transform of the
noise evaluated at $t=0$, \ie\ the autocorrelation for no time
difference. The noise before filtering is given by equation~\ref{eq:genNin}.
The mean square noise is
\begin{equation}
n=\frac1{\pi}\int_0^{\infty}{N(\w)|F(\w)|^2\,d\w} .
\end{equation}
The contribution of the electronic noise is
\begin{equation}
n_{elec}=\wc(\vn^2 C^2 \wc^2 + 3 \in^2)/32  .
\end{equation}
The peak of $\sqrt{n_{elec}/o^2(t)}$ gives the ENC for the electronic
noise after analog filtering. Appropriate scaling gives the ENE shown
in Figure~\ref{kev}.
The background noise contribution is
\begin{equation}
n_{back}=\frac{(8\wc^2+9\ws\wc+3\ws^2)\wc\ws\ib^2}{32(\wc+\ws)^3}
\end{equation}
$n$ is given by their sum. The signal to noise
ratio for a single sample taken at time $t$ is given by
$o^2(t)/n$. If we divide this by the optimum signal to noise given
by equation~\ref{eq:optsn} we get the ``efficiency'' for this
single sample versus the sample time. Figure~\ref{eff}
shows this function for the \babar\ calorimeter during
nominal operating conditions for several values of the corner frequency.
The highest efficiency achieved is 68\% for a shaping time
constant of 0.2\mus.
For low background conditions (Figure~\ref{effa})
the peak efficiency is 80\% at
a shaping time of 1.5\mus. At ten times nominal background,
(Figure~\ref{effb})
the efficiency drops to 50\% at a shaping time of 0.12\mus.
The loss of efficiency for nominal and better background
conditions is not significant. The loss at during high
background conditions hits just when noise is the causing
the most problems. Matched filters utilize knowledge of the
signal shape. In low background conditions, the optimum shaping
time is longer than the signal.
The shape of the signal is not seen and matched filters offer little
advantage. The situation and conclusions are reversed in high background
conditions.

The cost of using an analog filter optimized for conditions other
than those encountered is more dramatic. If the corner frequency
is set to the 0.2\mus\ which is optimal for nominal conditions, but
the backgrounds are actually at $10\times$ nominal levels, the efficiency is
44\%. If the corner frequency were optimal for low background conditions,
the efficiency at $10\times$ nominal background would be 12\%.

\section{Summary of Significant Findings}

Since the important conclusions may have gotten lost in the
derivations, I'll review them here.

\begin{itemize}

\item Machine background can be treated as a current noise
with the spectrum of the signal.

\item The best possible amplitude and time measurements can be obtained
with a matched filter.

\item The errors in the amplitude and time are not correlated, therefore
knowing the time does not improve the ultimate precision of the
amplitude determination, although such knowledge may reduce the
computation required.

\item The signal to noise methodology used here reproduces previous
results.

\item The background current noise dominates the photodiode shot noise
even at nominal backgrounds and in the quietest part of the calorimeter.

\item For nominal background where $\ib\gg\in$:
\begin{eqnarray*}
ENC^2&\approx&\ib^2\sqrt{2\ts\tb+\tb^2}
=\sqrt{2\ts\ib^3\vn\Ct+\ib^2\vn^2\Ct^2}\\
ENTC^2&\approx&ENC^2\ts\tb
=\sqrt{2\ib\ts^3\vn^3\Ct^3+\ts^2\vn^4\Ct^4}\\
\end{eqnarray*}
\item The sampling frequency of 3.7\mhz\ is adequate at
nominal backgraounds, but it is marginal
for $10\times$nominal background levels.

\item At this frequency, 64 samples will have to be processed
under nominal conditions.

\item The step of an ADC bit (on the most sensitive range)
should be set so that there are
at least 13 bits to full scale (12\gev) during nominal
background. For quiet conditions and
with a differentiating stage in the analog filter, 16.5 bits may
be required.

\item Simple analog filters can achieve near optimal results in
low to moderate background conditions if the shaping time is
optimized for the actual conditions.

\end{itemize}

\section{Acknowledgements}

I thank my BaBar colleagues for consultations and for providing the data
used in the examples. This work was supported by the U.S. Department of Energy 
under contract number DE-AC02-76SF00515.


\end{document}